\documentclass[twocolumn]{jpsj3} 
\usepackage{color}
\title{
Magnetization Jump in the Magnetization Process 
of the spin-1/2 Heisenberg Antiferromagnet 
on a Distorted Square-Kagome Lattice 
}
\catcode`\@=11
\def\simle{\mathrel{\mathpalette\@versim<}}   
\def\simge{\mathrel{\mathpalette\@versim>}}   
\def\@versim#1#2{\lower2.5pt\vbox{\baselineskip0pt \lineskip-.5pt
   \ialign{$\m@th#1\hfil##\hfil$\crcr#2\crcr\sim\crcr}}}
\catcode`\@=12

\author{Hiroki Nakano$^{1}$
\thanks{E-mail: hnakano@sci.u-hyogo.ac.jp} , 
Yasumasa Hasegawa$^{1}$
\thanks{E-mail: hasegawa@sci.u-hyogo.ac.jp} , 
and 
T\^oru Sakai$^{1,2}$
\thanks{E-mail: sakai@spring8.or.jp}
}

\inst{
$^{1}$Graduate School of Material Science, 
University of Hyogo,
Kamigori, 
Hyogo 678-1297, Japan \\
$^{2}$
Japan Atomic Energy Agency, SPring-8, 
Sayo, Hyogo 679-5148, Japan 
}

\recdate{\today}

\abst{
We study 
the magnetization process of the spin-$1/2$ Heisenberg 
antiferromagnet on a distorted square-kagome lattice 
by the numerical-diagonalization method.  
The magnetization jump at one-third of the height of the saturation 
is examined in detail; we find that 
the jump becomes larger when a small distortion is switched on 
and that it is accompanied 
by an abrupt change in lines along microscopic spin directions. 
Our finite-size results successfully confirm that 
the magnetization jump in a spin-isotropic system 
is a macroscopic jump 
that survives in the thermodynamic limit 
and that the changes in spin directions 
are common to a spin-flop phenomenon observed in spin-anisotropic systems. 
}

\begin{document}
\maketitle

\section{Introduction} 
Frustration has attracted much attention in condensed matter physics 
because it plays an important role as an origin of nontrivial 
behaviors in various systems. 
Particularly, such nontrivial behaviors are often observed 
in magnetic materials. 
One of them is the magnetization plateau. 
A magnetization plateau is the appearance 
of a region of a magnetic field in a magnetization process 
where the magnetization does not increase 
even with an increase in the magnetic field. 
This phenomenon is in contrast to a normal case, in which 
a magnetization process shows a smooth and significant
increase in magnetization with an increase in magnetic field. 
Such a plateau originates from the existence of 
an energy gap between levels in magnetic fields 
owing to the formation of an energetically stable quantum spin state, 
for example, the spin-1 alternating spin chain\cite{Narumi} and 
two-dimensional orthogonal dimer system\cite{Kageyama}. 

Another nontrivial behavior is the magnetization jump. 
During a jump in the magnetization process, 
the increase in magnetization is discontinuous, 
in contrast to the fact that a magnetization plateau corresponds 
to a discontinuity with respect to the magnetic field. 
As an origin of such magnetization jumps, 
a spin-flop phenomenon is well known\cite{Neel}. 
The phenomenon is the occurrence of an abrupt change 
in lines along microscopic spin directions 
while the states change owing to the increase in magnetic field. 
It is widely known that 
the phenomenon occurs when the system includes anisotropy 
in spin space\cite{kohno_MTakahashi_Maxwell_Construction,
TSakai_Maxwell_Construction}. 
Under these circumstances, there was a report 
of an interesting system that shows a magnetization jump 
even when the system has no anisotropy in spin space; 
the system is the spin-1/2 Heisenberg antiferromagnet 
on the square-kagome lattice\cite{shuriken_lett}. 

This lattice is originally introduced in the study of 
the relationship between the spin model on this
lattice and the eight-vertex model\cite{squagome_Siddharthan_Georges}; 
numerical-diagonalization studies 
of the quantum system on this lattice
were carried out\cite{squagome_Tomczak,squagome_Richter} 
although the above jump was not recognized within these studies. 
The lattice shares several characteristics with the kagome lattice; 
for example, the coordination number $z=4$ from a vertex of each lattice 
is the same. 
The corner sharing of neighboring local triangles is also the same. 
We hereafter call the structure of each unit cell a {\it shuriken} 
from its shape. 
The most important difference between the kagome and square-kagome lattices 
is whether or not all the vertices are equivalent 
with respect to the polygons that surround a focused vertex. 
In the square-kagome lattice, vertices are divided into two groups: 
one that is a vertex of the small square 
inside the local structure of a {\it shuriken} 
and one that is not (see Fig.~\ref{fig1}). 
Let us call a vertex site in the former (latter) group 
the $\alpha$ ($\beta$) site hereafter. 
This difference induces nontrivial different behaviors 
between the magnetization processes of these two systems. 
In the kagome-lattice antiferromagnet, there is a region 
similar to a magnetization plateau at one-third height of the saturation. 
Just outside of this height, the change in the magnetization is 
continuous but the magnetization process shows 
a nontrivial critical behavior, which is certainly different 
from a magnetization plateau of the well-known type 
in a two-dimensional system\cite{kgm_ramp,Sakai_HN_PRBR,HN_TSakai_kgm_1_3}. 
Although there appears a magnetization plateau 
at the same height in the square-kagome-lattice antiferromagnet, 
the plateau was found to be accompanied by a discontinuous change 
in the magnetization at the edge of the higher-field side 
in the magnetization processes of finite-size clusters. 
The discontinuity strongly suggests a magnetization jump. 
The local magnetization also shows its discontinuity, 
which suggests that a behavior similar to a spin-flop phenomenon happens. 
A common behavior in finite-size clusters is observed in 
other cases: 
the Cairo-pentagon lattice\cite{HNakano_Cairo_lt,Isoda_Cairo_full}, 
the distorted kagome lattice\cite{HN_kgm_dist,HN_TSakai_kgm_1_3}, and 
the {\it shuriken}-bonded honeycomb lattice\cite{HN_kgm_dist}. 

However, to confirm that 
this discontinuous behavior of finite-size clusters 
in the square-kagome-lattice antiferromagnet 
corresponds to a true magnetization jump owing to a spin-flop phenomenon, 
the following two viewpoints should be clarified. 
One is whether or not the jump certainly survives 
in the thermodynamic limit. 
In a normal magnetization process without any anomalies, 
the increase in magnetization of a finite-size cluster 
must be $\delta M = 1$; 
in the square-kagome-lattice antiferromagnet, 
a discontinuity of $\delta M = 2$ was found. 
Note here that a discontinuity of $\delta M = 2$ can appear 
in the spin-nematic state\cite{nematic_momoi,nematic_shanonn}. 
Since the spin-nematic state is the two-spin magnon state, 
the discontinuity of the spin-nematic state is not a macroscopic behavior. 
To exclude the possibility that a discontinuity of $\delta M = 2$ 
in the square-kagome-lattice antiferromagnet indicates the spin-nematic state, 
it is sufficient to observe a discontinuity of $\delta M \ge 3$. 
The other point is how the transverse spin component behaves. 
The local magnetization examined in Ref.~\ref{shuriken_lett} provides 
information on a component along the external field, namely, the $z$-axis. 
If the transverse spin component between both sides 
at the observed jump shows a significant difference, 
the relationship between the magnetization jump and the spin-flop phenomenon 
is established. 
Under these circumstances, 
the purpose of this study is to clarify 
these two unresolved issues in the square-kagome-lattice antiferromagnet 
by taking distortion into account additionally 
to the undistorted square-kagome-lattice case.  

Candidate square-kagome-lattice antiferromagnet materials 
have not been reported so far to the best of our knowledge. 
On the other hand, 
the discoveries of candidate kagome-lattice antiferromagnet materials 
such as 
herbertsmithite\cite{Shores_herbertsmithite2005,Mendels_herbertsmithite2010}, 
volborthite\cite{Yoshida_jpsj_volborthite2009,
Yoshida_prl_volborthite2009,Ishikawa_PRL_volborthite2015}, and 
vesignieite\cite{Okamoto_jpsj_vesignieite2009,
Okamoto_prb_vesignieite2011},  
have accelerated experimental studies 
as well as theoretical ones\cite{Lecheminant,Waldtmann,Hida_kagome2,
cabra2002,Honecker0,Honecker1,cabra2005,Cepas,
Jiang2008,Spin_gap_Sindzingre,2d_ferri,kgm_gap,Honecker2011,
Yan_Huse_White_DMRG,Depenbrock2012,capponi2013,
SNishimoto_NShibata_CHotta,HN_Sakai_PSS,HN_Sakai_SCES,Iqbal2014,Iqbal2015}, 
leading to a deeper understanding 
of the physics of the kagome-lattice antiferromagnet. 
From these studies, 
it is gradually clarified that the materials have 
significant deviations from the ideal situation 
of the kagome-lattice Heisenberg antiferromagnet. 
In volborthite, for example, magnetic interactions deviate from 
the uniform situation in the kagome lattice 
because of the orbital directions
despite the fact that atomic positions are located 
at the vertices of the kagome lattice. 
To understand the candidate materials well, therefore, 
effects due to the deviation from the ideal situation 
are worth studying, which is a background of this study. 

This paper is organized as follows. 
In the next section, the model that we study here is introduced. 
The method and analysis procedure are also explained. 
The third section is devoted to the presentation and 
discussion of our results. 
We first discuss the change in local magnetizations of the state 
at one-third of the height of the saturation. 
The discussion will clarify which phase 
the undistorted square-kagome-lattice case is located in. 
After that, the magnetization process of larger clusters is presented. 
The spin-spin correlations are also examined. 
In the final section, 
we present our conclusion. 

\section{Model Hamiltonians and Method} 

\begin{figure}[tb]
\begin{center}
\includegraphics[width=8cm]{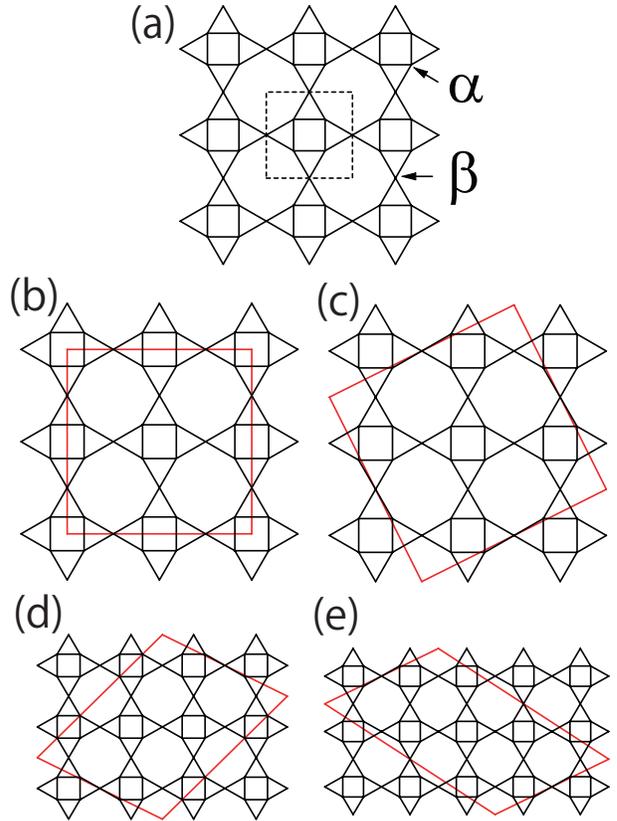}
\end{center}
\caption{(Color) 
Structure of the square-kagome lattice. 
In panel (a), $\alpha$ and $\beta$ sites are illustrated;  
the broken-line square represents a unit cell of this lattice. 
The shapes of finite-size clusters studied in this work 
are shown 
for (b) $N_{\rm s}=24$, (c) $N_{\rm s}=30$, (d) $N_{\rm s}=36$, 
and (e)  $N_{\rm s}=42$. 
}
\label{fig1}
\end{figure}
The Hamiltonian that we study in this work is given by 
${\cal H}={\cal H}_0 + {\cal H}_{\rm Zeeman}$, where 
\begin{equation}
{\cal H}_0 = \sum_{\langle i,j\rangle, i \in \alpha, j \in \beta} J_{1} 
\mbox{\boldmath $S$}_{i}\cdot\mbox{\boldmath $S$}_{j}
+ \sum_{\langle i,j\rangle, i \in \alpha, j \in \alpha} J_{2}
\mbox{\boldmath $S$}_{i}\cdot\mbox{\boldmath $S$}_{j}
 , 
\label{H_square_kagome}
\end{equation}
and 
\begin{equation}
{\cal H}_{\rm Zeeman} = - h \sum_{j} S_{j}^{z} .  
\label{H_zeeman}
\end{equation}
Here, $\mbox{\boldmath $S$}_{i}$ 
denotes the $S=1/2$ spin operator 
at site $i$, where the sites are the vertices of the square-kagome lattice 
shown in Fig.~\ref{fig1}. 
The spin operator satisfies $\mbox{\boldmath $S$}_{i}^2=S(S+1)$. 
The sum of ${\cal H}_0$ runs over all the nearest-neighbor pairs 
in the square-kagome lattice. 
Energies are measured in units of $J_{1}$; 
hereafter, we set $J_{1}=1$.  
The number of spin sites is denoted by $N_{\rm s}$, 
where $N_{\rm s}/6$ is an integer. 
We treat $N_{\rm s}=24$, 30, 36, and 42 in this study. 
We impose the periodic boundary conditions 
for clusters with site $N_{\rm s}$; 
the shapes of the clusters are shown in Fig.~\ref{fig1}.  
Note here that the shapes for $N_{\rm s}=24$ and 30 are squares 
although the latter is slightly tilted. 
On the other hand, the shapes for $N_{\rm s}=36$ and 42 
are not squares. 
However, examination of systems with a large $N_{\rm s}$ would 
give a higher resolution of the magnetization process and 
contribute much to our understanding of this system. 
Note here that the same model including the control of $J_{2}$ 
was previously examined in Refs.~\ref{Derzhko_Richter_PRB}
and \ref{Rousochatzakis_Moessner_Brink_PRB}.  
Reference~\ref{Derzhko_Richter_PRB} treated this model 
within an effective Hamiltonian based on the localized-magnon picture. 
Reference~\ref{Rousochatzakis_Moessner_Brink_PRB} treated 
only $N_{\rm s}=24$ and 30. 
Both of these studies did not examine the magnetization jump 
at one-third of the height of the saturation on which we focus our attention 
in this study. 

We calculate the lowest energy of ${\cal H}_0$ 
in the subspace belonging to $\sum _j S_j^z=M$ 
by numerical diagonalizations 
based on the Lanczos algorithm and/or Householder algorithm. 
The energy is denoted by $E(N_{\rm s},M)$, 
where $M$ takes an integer value 
down from $M_{\rm low}$ (=0) 
up to the saturation value $M_{\rm s}$ ($=S N_{\rm s}$). 
We use the normalized magnetization $m=M/M_{\rm s}$. 
Some of the Lanczos diagonalizations were carried out 
using the MPI-parallelized code, which was originally 
developed in the study of Haldane gaps\cite{HN_Terai}. 
The usefulness of our program was previously confirmed in large-scale 
parallelized calculations\cite{kgm_gap,s1tri_LRO,HN_TSakai_kgm_1_3}. 
Note here that the calculations in cases of a small $M$ for $N_{\rm s}=42$ 
require the use of the K computer (Kobe, Japan) 
to obtain the magnetization process of the entire range. 

The magnetization process for a finite-size system is obtained 
by considering the magnetization increase from $M$ to $M+1$ in the field, 
\begin{equation}
H=E(N_{\rm s},M+1)-E(N_{\rm s},M),
\label{field_at_M}
\end{equation}
under the condition that the lowest-energy state 
with the magnetization $M$ and that with the magnetization $M+1$ 
become the ground state in specific magnetic fields.  
It is often the case that the lowest-energy state with the magnetization $M$ 
does not become the ground state in any field. 
In this case, the magnetization process around the magnetization $M$ 
is determined by the Maxwell 
construction\cite{kohno_MTakahashi_Maxwell_Construction,
TSakai_Maxwell_Construction}. 

We evaluate the average local magnetization defined as
\begin{equation}
m_{\rm LM}^{\xi}=\frac{1}{N_{\xi}} 
\sum_{j\in \xi} \langle S_{j}^{z}\rangle , 
\label{local_m}
\end{equation}
where $\xi$ takes $\alpha$ and $\beta$. 
$N_{\xi}$ denotes the number of $\xi$ sites; 
namely, $N_{\alpha}=2N_{\rm s}/3$ and $N_{\beta}=N_{\rm s}/3$.  
Here, the symbol $\langle {\cal O} \rangle$ represents 
the expectation value of the operator ${\cal O}$ 
with respect to the lowest-energy state 
within the subspace characterized by a fixed magnetization $M$ 
that we focus our attention on. 
Averaging over $\xi$ is carried out 
when the ground-state level is degenerate. 
Note that, when the ground state belonging to $M$ is not degenerate, 
the results do not change 
regardless of the presence or absence of this average. 
We also measure the correlation functions 
$\langle S_{j}^{z} S_{k}^{z} \rangle$ and $\langle S_{j}^{x} S_{k}^{x} \rangle$ 
to capture well 
the features of the wave functions that are numerically obtained. 

\section{Results and Discussion} 

\begin{figure}[tb]
\begin{center}
\includegraphics[width=8cm]{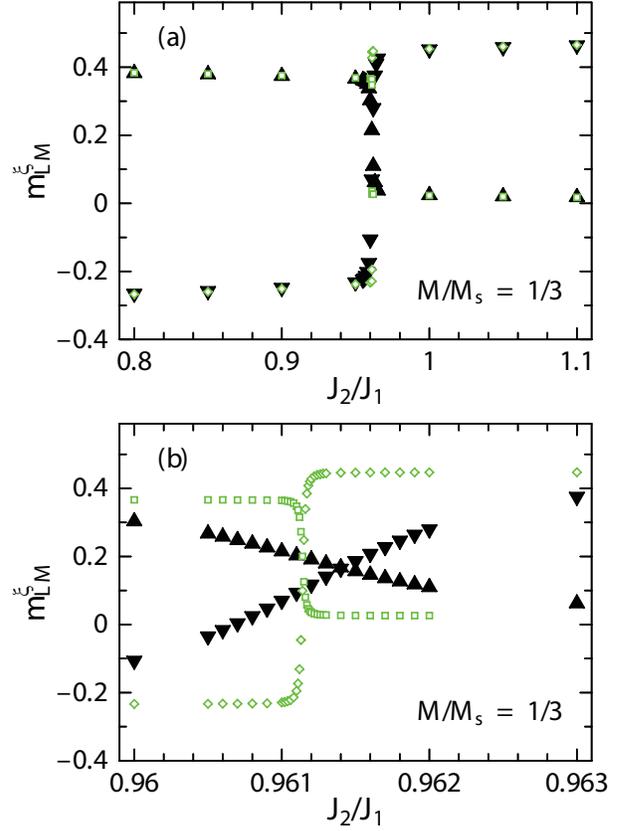}
\end{center}
\caption{(Color) 
Dependence of the local magnetization at $m=1/3$ 
on the ratio of $J_2/J_1$.  
Panel (b) is a zoom-in view of panel (a) 
to observe the range near the critical point in detail. 
Black closed triangles and inversed triangles denote 
$N_{\rm s}=24$ data of $\alpha$ and $\beta$, respectively. 
Green open squares and diamonds denote 
$N_{\rm s}=30$ data of $\alpha$ and $\beta$, respectively. 
}
\label{fig2}
\end{figure}
First, we present our results of 
the $J_{2}$ dependence of the local magnetization of the $m=1/3$ state, 
which are shown in Fig.~\ref{fig2}. 
A similar examination concerning the parameter dependence 
was carried out in cases of the Cairo-pentagon 
lattice\cite{HNakano_Cairo_lt} 
and the distorted kagome lattice\cite{HN_kgm_dist}, and 
the Lieb lattice accompanied 
by frustrating interaction\cite{HN_TSakai_JJAP_RC}. 
One finds in Fig.~\ref{fig2}(a) 
a large decrease in $m_{\rm LM}^{\alpha}$ 
and a large increase in $m_{\rm LM}^{\beta}$ near $J_{2}/J_{1}\sim 0.96$. 
This marked change suggests that 
a phase transition occurs 
between the ferrimagnetic state for $J_{2}/J_{1} \simle 0.96$ 
and the state composed of singlet spins and almost fully polarized spins 
for $J_{2}/J_{1} \simge 0.96$.  
The phase transition has already been reported 
in Ref.~\ref{Rousochatzakis_Moessner_Brink_PRB}, 
which concluded that this is a first-order transition. 
To detect the behavior of $m_{\rm LM}^{\xi}$ near $J_{2}/J_{1}\sim$0.96, 
we observe a detailed change in the zoom-in view in Fig.~\ref{fig2}(b). 
Our results show a continuous change in $m_{\rm LM}^{\xi}$ 
for both $N_{\rm s}=24$ and 30 
although the change for $N_{\rm s}=30$ is much more rapid 
than that for $N_{\rm s}=24$. 
From the observation of this continuous behavior, 
the conclusion of the first-order transition may be premature. 
Note also that the continuous change for $N_{\rm s}=30$ is 
in contrast to the discontinuous change for the same $N_{\rm s}$ 
in the case of the Cairo-pentagon lattice\cite{HNakano_Cairo_lt} 
despite the fact that 
the present model and the model on the Cairo-pentagon lattice 
share the same number of spin sites in each unit cell. 
It is difficult within studies based on finite-size examinations, 
but it should be resolved in future studies 
to determine whether the transition is of the second or first order.  
An important point clarified in this study is 
that $J_{2}/J_{1}=1$ is different from the transition point $\sim$ 0.96 
and 
that $J_{2}/J_{1}=1$ is in the state 
composed of singlet spins in the $\alpha$ sites 
and 
almost fully polarized spins in the $\beta$ sites. 
Hereafter, in the present study, we focus on the case of the side 
of $J_{2}/J_{1}$ that is greater than the transition point 
$J_{2}/J_{1} \sim 0.96$.  


\begin{figure}[tb]
\begin{center}
\includegraphics[width=8cm]{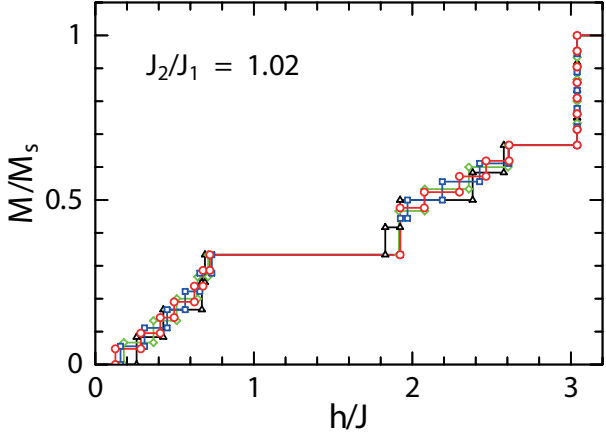}
\end{center}
\caption{(Color) 
Magnetization process for $J_2/J_1=1.02$. 
Black triangles, green diamonds, blue squares, and red circles 
denotes data for $N_{\rm s}=24$, 30, 36, and 42, respectively. 

}
\label{fig3}
\end{figure}
\begin{figure}[tb]
\begin{center}
\includegraphics[width=8cm]{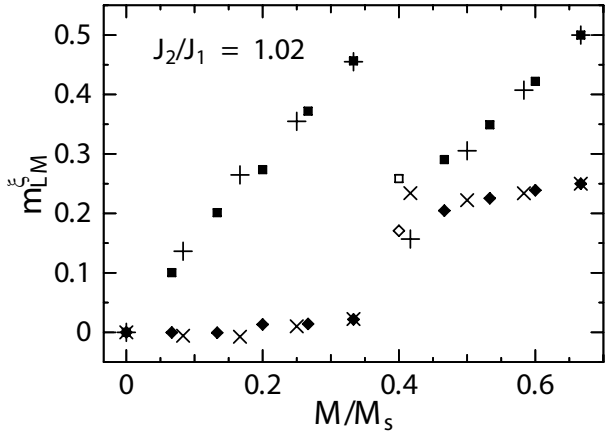}
\end{center}
\caption{
Average local magnetization at $m=1.02$. 
Pluses and crosses denote $N_{\rm s}=24$ data for $\alpha$ and $\beta$, 
respectively. 
Diamonds and squares denote $N_{\rm s}=30$ data for $\alpha$ and $\beta$, 
respectively. 
All the states are realized for $N_{\rm s}=24$, but not for $N_{\rm s}=30$.  
For $N_{\rm s}=30$, closed symbols represent data 
for the stably realized states, 
while open symbols denote data for the unstable states 
during the magnetization jump.
}
\label{fig4}
\end{figure}

Next, we examine the magnetization process 
in the phase of the $J_{2}/J_{1}=1$ side. 
Our results are shown in Fig.~\ref{fig3} for $J_{2}/J_{1}=1.02$. 
This parameter is chosen within the same side as $J_{2}/J_{1}=1$ 
as a value close to $J_{2}/J_{1}=1$. 
Some features are shared between $J_{2}/J_{1}=1$ and 1.02; 
a detailed comparison between the results of $J_{2}/J_{1}=1$ and 1.02 
will be carried out later. 
The magnetization processes in Fig.~\ref{fig3} show 
a clear existence of the $m=1/3$ plateau. 
No jump appears for $N_{\rm s}=24$ 
at the edge of the higher-field side of the $m=1/3$ height. 
For $N_{\rm s} \ge 30$, on the other hand, 
magnetization jumps are clearly observed. 
The appearance of such jumps 
is independent of $N_{\rm s}$,  
at least within $N_{\rm s} \ge 30$ of the present finite-size-system study. 
In Fig.~\ref{fig4}, the average local magnetizations 
for the square-cluster systems are shown.  
One easily finds that discontinuous behaviors 
in $\alpha$ and $\beta$ sites appear 
between $M=(1/3)M_{\rm s}$ and $M> (1/3)M_{\rm s}$. 
At $M=(1/3)M_{\rm s}$, our results 
suggest that 
a spin at a $\beta$ site reveals almost a full moment along the $z$-axis 
and that 
spins at $\alpha$ sites form a singlet state in the small square 
inside each {\it shuriken}. 

\begin{figure}[tb]
\begin{center}
\includegraphics[width=8cm]{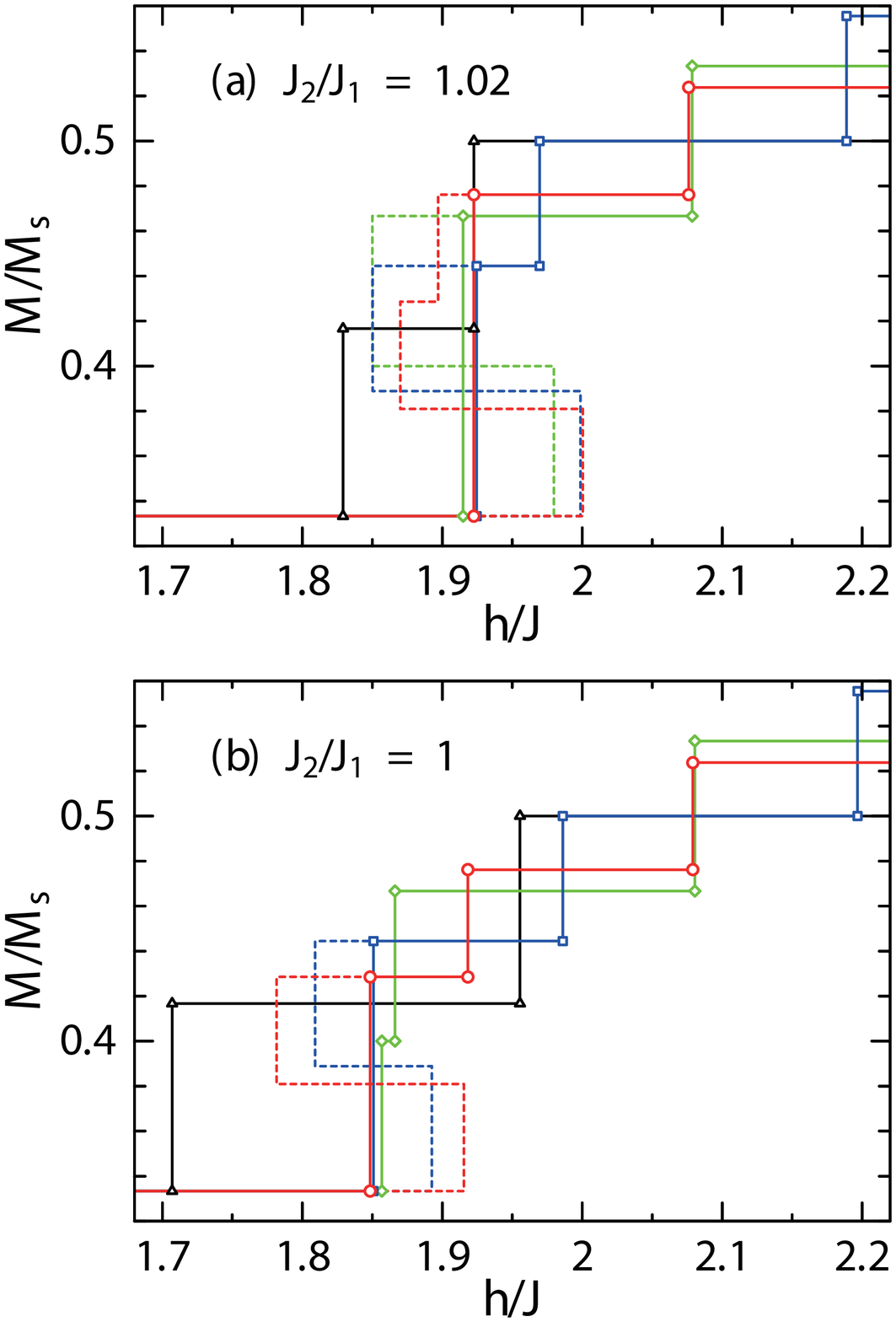}
\end{center}
\caption{(Color) 
Zoom-in views of the magnetization processes for (a) $J_2/J_1=1.02$ 
and (b) $J_2/J_1=1$. 
Symbols are the same as those in Fig.~\ref{fig3}. 
The broken lines represent the results 
before the Maxwell construction is carried out. 
}
\label{fig5}
\end{figure}
Let us next compare between the results of the magnetization processes 
for $J_{2}/J_{1}=1$ and 1.02. 
Figures~\ref{fig5}(a) and \ref{fig5}(b) show zoom-in views 
of the magnetization processes 
around the edge of the higher-field side of the $m=1/3$ height. 
The most marked difference between the two cases 
is the skip $\delta M$ in the results for $N_{\rm s}=42$; 
we observe $\delta M = 2$ for $J_{2}/J_{1}=1$ whereas 
we observe $\delta M = 3$ for $J_{2}/J_{1}=$1.02. 
The result of $\delta M = 3$ for $J_{2}/J_{1}=$1.02 is important 
from the viewpoint of the origin of the jump. 
A finite-size jump with $\delta M = 2$ can appear 
in the nematic phase\cite{nematic_momoi,nematic_shanonn}, 
where the two-magnon bound state is realized. 
The nematic phase is gapless so that 
the magnetization process in the thermodynamic limit is continuous. 
Thus, the finite-size jump owing to the nematic phase 
is only a finite-size effect and 
does not become a macroscopic jump in the thermodynamic limit. 
The present result of $\delta M = 3$ for $N_{\rm s}=42$ and $J_{2}/J_{1}=$1.02 
clearly excludes the possibility for the appearance of the finite-size jump 
owing to the nematic phase. 
The skip $\delta M = 3$ also suggests that 
the jump becomes a macroscopic one in the thermodynamic limit. 

Let us then estimate the skip in the thermodynamic limit 
from our finite-size results.  
The resolution of the normalized magnetization process 
is given by $2/N_{\rm s}$. 
If a finite-size system reveals a jump with $\delta M =2$, 
the skip $\delta m$ of the normalized magnetization jump 
in the thermodynamic limit should satisfy 
\begin{equation} 
2/N_{\rm s} < \delta m < 2 (2/N_{\rm s}), 
\label{one_skip}
\end{equation} 
while the skip is given by 
\begin{equation} 
 \delta m  > 2(2/N_{\rm s}) , 
\label{two_skip}
\end{equation} 
if a finite-size system reveals a jump with $\delta M =3$.  
For $J_{2}/J_{1}=$1.02, 
we have $\delta M = 2$ for $N_{\rm s}=36$ and $\delta M = 3$ for $N_{\rm s}=42$; 
Eqs.~(\ref{one_skip}) and (\ref{two_skip}) suggest 
\begin{equation}
0.095 \simle \delta m \simle 0.111 . 
\label{skip_1_02}
\end{equation} 
For $J_{2}/J_{1}=$1, on the other hand, 
we have $\delta M=2$ for $N_{\rm s}=36$ and 42 and no jumps 
for $N_{\rm s}\le 30$, 
which suggest 
\begin{equation}
\delta m \simle 0.095  . 
\label{skip_1}
\end{equation} 
Therefore, one finds that 
the skip of the jump becomes larger when $J_{2}/J_{1}$ 
deviates from the critical point $\sim 0.96$ clarified 
in Fig.~\ref{fig2}.  


\begin{table}[hbt]
\caption{
Correlation functions for $N_{\rm s}=24$ cluster 
at $J_{2}/J_{1}= 1.02$. 
We select a pair of sites $i$ and $j$ so that 
the distance between these sites is largest in this cluster. }
\label{table1}
\begin{center}
\begin{tabular}{ccccc}
\hline
 &     & $M=\frac{1}{3}M_{\rm s}$ & $M=\frac{1}{3}M_{\rm s}+1$ 
 & $M=\frac{1}{3}M_{\rm s}+2$ \\
\hline
$i,j\in \alpha$  
  & \multicolumn{1}{@{}c@{}}{$\begin{array}{c}
                             \langle S_{i}^{z} S_{j}^{z}\rangle  \\
                             \langle S_{i}^{x} S_{j}^{x}\rangle  \end{array}$}
  & \multicolumn{1}{@{}c@{}}{$\begin{array}{r}
                             0.0005068   \\
                             0.0000012   \end{array}$}
  & \multicolumn{1}{@{}c@{}}{$\begin{array}{c}
                             0.0571670  \\
                             0.0048659  \end{array}$}
  & \multicolumn{1}{@{}c@{}}{$\begin{array}{c}
                             0.0498709  \\
                             0.0043312  \end{array}$}\\
\hline
$i,j\in \beta$ 
  & \multicolumn{1}{@{}c@{}}{$\begin{array}{c}
                             \langle S_{i}^{z} S_{j}^{z}\rangle  \\
                             \langle S_{i}^{x} S_{j}^{x}\rangle  \end{array}$}
  & \multicolumn{1}{@{}c@{}}{$\begin{array}{r}
                             0.2071796  \\
                             0.0000184  \end{array}$}
  & \multicolumn{1}{@{}c@{}}{$\begin{array}{c}
                             0.0335379  \\
                             0.0192386  \end{array}$}
  & \multicolumn{1}{@{}c@{}}{$\begin{array}{c}
                             0.0960927  \\
                             0.0167820  \end{array}$}\\
\hline
\end{tabular}
\end{center}
\end{table}
Next, let us consider the spin direction from the data of 
correlation functions. 
We measure 
$\langle S_{i}^{z} S_{j}^{z}\rangle$ and $\langle S_{i}^{x} S_{j}^{x}\rangle$ 
between the pair of sites $i$ and $j$, where 
$i$ and $j$ in the same group are the most distant in a finite-size cluster 
forming a square. 
The same analysis was carried out in Ref.~\ref{HN_kgm_dist}. 
Within this study, the $N_{\rm s}=24$ cluster is 
the case that should be measured 
while the others are not 
because the $N_{\rm s}=42$ and 36 clusters do not form a square 
and 
because 
$i$ 
and $j$ as the most distant sites for the $N_{\rm s}=30$ cluster 
are not in the same group.   
Our present results are summarized for $J_{2}/J_{1}=1.02$ 
in Table~\ref{table1}. 
Note here that, for $N_{\rm s}=24$, 
$M=(1/3)M_{\rm s}$, $M=(1/3)M_{\rm s}+1$, and $M=(1/3)M_{\rm s}+2$ 
correspond to $m=1/3$, $m=0.4167$, and $m=0.5$, respectively. 
From Eq.~(\ref{skip_1_02}), 
it is clear that $m=0.5$ is outside the jump, 
although 
it is unclear whether $m=0.4167$ is outside or inside the jump 
because $m=0.4167$ is very close to the upper edge of the jump. 
Therefore, 
it is reasonable to consider that 
the $M=(1/3)M_{\rm s}+2$ state is appropriate for observing 
a state under a field higher than the jump; 
on the other hand, 
the $M=(1/3)M_{\rm s}$ state 
is appropriate for observing 
a state under a field lower than the jump. 
At $M=(1/3)M_{\rm s}$, only $\langle S_{i}^{z} S_{j}^{z}\rangle$ 
for $i,j \in \beta$ is dominant; 
the other quantities are very small. 
These results are in agreement with the behavior that we have observed 
in the $m_{\rm LM}^{\xi}$. 
Thus, it is confirmed that, at $m=1/3$, the system forms the state 
in which 
the spins at $\beta$ sites have almost a full moment along the $z$-axis 
and where 
spins at $\alpha$ sites form a singlet state in the small square 
inside each {\it shuriken}. 
At $M=(1/3)M_{\rm s}+2$, on the other hand, 
the spin state is markedly different. 
$\langle S_{i}^{z} S_{j}^{z}\rangle$ for $i,j \in \beta$ 
is markedly smaller than that at $M=(1/3)M_{\rm s}$ 
while 
$\langle S_{i}^{z} S_{j}^{z}\rangle$ for $i,j \in \alpha$ is larger. 
In addition, both 
$\langle S_{i}^{x} S_{j}^{x}\rangle$ for $i,j \in \alpha$ 
and 
$\langle S_{i}^{x} S_{j}^{x}\rangle$ for $i,j \in \beta$ 
are significantly nonzero. 
The results concerning the transverse component strongly suggest that 
$\alpha$ and $\beta$ spins take directions that have nonzero angles 
from the $z$-axis. 
This picture concerning the state of $M=(1/3)M_{\rm s}+2$ 
is consistent with the speculation for $J_{2}/J_{1}=1$ 
presented in Ref.~\ref{shuriken_lett} as the umbrella-type spin state.  
Finally, note that 
the behavior at $M=(1/3)M_{\rm s}+1$ is qualitatively the same 
as that at $M=(1/3)M_{\rm s}+2$, 
although it is unclear 
whether the $m=0.4167$ state is stable or unstable. 

\section{Conclusion and Remarks} 

In this paper, we study the spin-1/2 Heisenberg antiferromagnet 
on a square-kagome lattice with distortion 
by numerical-diagonalization calculations. 
We have found that the magnetization jump occurs 
at the higher-field edge of one-third of the height of the saturation 
as a frustration effect.  
An essential improvement from the results 
in Ref.~\ref{shuriken_lett} is that 
our large-scale calculations detect the behavior of the growth of the jump 
when the interactions in the small square in a {\it shuriken} become larger.  
Our finite-size results have clarified that 
a possibility of a finite-size jump owing to the spin-nematic state 
is excluded. 
Our analysis of spin-spin correlation functions 
is another essential improvement suggesting 
that, in the higher-field region, each spin reveals 
its own significant transverse component,  
while no transverse components are observed 
in the state at one-third of the height of the saturation. 
The changes in spin directions in the spin-isotropic system 
share the behavior in the spin-flop phenomenon observed 
in spin-anisotropic systems. 
Magnetization jumps in spin-isotropic systems 
were reported in several studies\cite{Fouet_PRB2006,Michaud_PRB2010,
Honecker_Mila_Troyer,Schroder_icosahedron,
Konstantinidis_icosahedron,TMomoi_KTotsuka,KKubo_TMomoi,
TMomoi_HSakamoto_KKen,KPenc_NShannon_HShiba}.  
The comparison between these cases and the present case 
will also be useful for understanding well the occurrence of these jumps. 
The present consequence of the magnetization jump 
will become a clue for future investigations, which 
would contribute much to 
our understanding of frustration effects in magnetic materials. 

\section*{Acknowledgments}
This work was partly supported 
by JSPS KAKENHI Grant Numbers 23340109 and 24540348. 
Nonhybrid thread-parallel calculations in
numerical diagonalizations were based on TITPACK version
2 coded by H. Nishimori. 
This research used computational resources of the K computer 
provided by the RIKEN Advanced Institute for Computational Science 
through the HPCI System Research projects (Project ID: hp150024). 
Some of the computations were 
performed using facilities of 
the Department of Simulation Science, 
National Institute for Fusion Science; 
Center for Computational Materials Science, 
Institute for Materials Research, Tohoku University; 
Supercomputer Center, 
Institute for Solid State Physics, The University of Tokyo;  
and Supercomputing Division, 
Information Technology Center, The University of Tokyo. 
This work was partly supported 
by the Strategic Programs for Innovative Research; 
the Ministry of Education, Culture, Sports, Science 
and Technology of Japan; 
and the Computational Materials Science Initiative, Japan. 
We would like to express our sincere thanks 
to the staff of the Center for Computational Materials Science 
of the Institute for Materials Research, Tohoku University, 
for their continuous support 
of the SR16000 supercomputing facilities. 


\end{document}